\documentclass[journal=jacsat,manuscript=communication,layout=twocolumn]{achemso}

\usepackage{graphicx}
\usepackage{physics}
\usepackage{siunitx}
\usepackage{amssymb}
\usepackage{soul}

\SectionsOn
\mciteErrorOnUnknownfalse

\title{Gatemon qubit on a germanium quantum-well heterostructure}

\author{Elyjah Kiyooka}
\affiliation{Univ. Grenoble Alpes, CEA, Grenoble INP, IRIG, PHELIQS, 38000 Grenoble, France}
\email{elyjah.kiyooka@gmail.com}
\author{Chotivut Tangchingchai}
\affiliation{Univ. Grenoble Alpes, CEA, Grenoble INP, IRIG, PHELIQS, 38000 Grenoble, France}
\author{Leo Noirot}
\affiliation{Univ. Grenoble Alpes, CEA, Grenoble INP, IRIG, PHELIQS, 38000 Grenoble, France}
\author{Axel Leblanc}
\affiliation{Univ. Grenoble Alpes, CEA, Grenoble INP, IRIG, PHELIQS, 38000 Grenoble, France}
\author{Boris Brun}
\affiliation{Univ. Grenoble Alpes, CEA, Grenoble INP, IRIG, PHELIQS, 38000 Grenoble, France}
\author{Simon Zihlmann}
\affiliation{Univ. Grenoble Alpes, CEA, Grenoble INP, IRIG, PHELIQS, 38000 Grenoble, France}
\author{Romain Maurand}
\affiliation{Univ. Grenoble Alpes, CEA, Grenoble INP, IRIG, PHELIQS, 38000 Grenoble, France}
\author{Vivien Schmitt}
\affiliation{Univ. Grenoble Alpes, CEA, Grenoble INP, IRIG, PHELIQS, 38000 Grenoble, France}
\author{\'{E}tienne Dumur}
\affiliation{Univ. Grenoble Alpes, CEA, Grenoble INP, IRIG, PHELIQS, 38000 Grenoble, France}
\author{Jean-Michel Hartmann}
\affiliation{Univ. Grenoble Alpes, CEA, LETI, 38000 Grenoble, France}
\author{Francois Lefloch}
\affiliation{Univ. Grenoble Alpes, CEA, Grenoble INP, IRIG, PHELIQS, 38000 Grenoble, France}
\author{Silvano De Franceschi}
\affiliation{Univ. Grenoble Alpes, CEA, Grenoble INP, IRIG, PHELIQS, 38000 Grenoble, France}
\email{silvano.defranceschi@cea.fr}


\begin{document}

\begin{abstract}
Gatemons are superconducting qubits resembling transmons, with a gate-tunable semiconducting weak link as the Josephson element. Here, we report a gatemon device featuring an aluminum microwave circuit on a Ge/SiGe heterostructure embedding a Ge quantum well. Owing to the superconducting proximity effect, the high-mobility two-dimensional hole gas confined in this well provides a gate-tunable superconducting  weak link between two Al contacts. We perform Rabi oscillation and Ramsey interference measurements, demonstrate the gate-voltage dependence of the qubit frequency, and measure the qubit anharmonicity. We find relaxation times $T_{1}$ up to \SI{119}{\nano\s}, and Ramsey coherence times $T_{2}^{*}$ up to \SI{70}{\nano\s}, and a qubit frequency gate-tunable over \SI{3.5}{\giga\hertz}. The reported proof-of-concept reproduces the results of a very recent work [Sagi et al., Nat. Commun. 15, 6400 (2024)] using similar Ge/SiGe heterostructures thereby validating a novel platform for the development of gatemons and parity-protected $\cos(2\phi)$ qubits.
\end{abstract}

\section{Introduction}

Superconducting transmon qubits based on traditional Al/AlO$_{x}$ Josephson junctions form today one of the most advanced physical platforms for  quantum computing \cite{DavideCastelvecchi2023,Arute2019}.  An alternative version of these qubits uses a weak-link made of semiconductor material so that the critical current, and thus the qubit energy, can be controlled by changing the carrier density in the weak link through an electrostatic gate voltage. This special type of transmon is known as gate-tunable transmon, or gatemon for short. So far, it has been realized using various materials such as InAs nanostructures \cite{Larsen2015,DeLange2015,Casparis2018,Strickland2024}, graphene \cite{Wang2019}, carbon nanotubes \cite{Mergenthaler2021}, core/shell Ge/Si one-dimensional nanowires \cite{Zheng2024,Zhuo2023}, and, only recently, Ge/SiGe two-dimensional (2D) nanostructures \cite{Sagi2024} similar to those used here, which have emerged as a novel versatile platform for both spin qubits and superconductor/semiconductor devices\cite{Scappucci2021}

In most of the superconducting qubit architectures developed so far, the energies of the transmon qubits, as well as their reciprocal couplings, are controlled via magnetic fluxes generated by locally circulating currents. This approach may become technically challenging at very large scale due to energy dissipation. Owing to their electrostatic control gatemons could offer a way around this potential problem. They could form voltage-tunable coupling elements and, provided their coherence and relaxation times can be sufficiently improved, they could even replace the transmons themselves.  

Additionally, the nature of the weak link for superconducting-semiconducting devices is different from the traditional Al/AlO$_{x}$ tunnel junctions. In particular, semiconducting weak links \cite{Vigneau2019} can have conduction channels with high transparency ($\tau \rightarrow$ 1)\cite{Tosato2023, Kjaergaard2017, Aggarwal2021, Leblanc2023} and ballistic charge carriers leading to a modified Josephson junction potential \cite{Beenakker1991, Kringhøj2018} and to the appearance higher harmonics in the current phase relation \cite{Leblanc2023,Banszerus2024}. 

Importantly, for a gatemon, a change in the Josephson potential manifests itself as a reduced energy difference between consecutive energy levels called the anharmonicity ($\alpha$) \cite{Kringhøj2018}. This reduction of the anharmonicity, already observed in many gatemon measurements  \cite{Kringhøj2018,Zheng2024,Sagi2024}, can then be used as a method for studying the fundamental properties of high transparency modes on a Josephson junction potential. 

Additionally, a current phase relation with higher harmonics can be used to create what is referred to as a $\cos{(2\phi)}$ qubit by, in a squid geometry, having the $\cos{(\phi)}$ component canceled between the two squid arms leaving mainly a $\cos{(2\phi)}$ component \cite{Larsen2020}. When used as a circuit element shunted with a capacitance, the resulting $\pi$ periodic potential has two minima creating a parity protection to relaxation \cite{Doucot2012, Leblanc2024}.

Here, we present a gatemon qubit made out of a Ge/SiGe heterostructure embedding a 2D hole gas characterized in a circuit quantum electrodynamics (cQED) experiment. The design and fabrication are discussed along with the characterization of the gate-dependent gatemon characteristics, such as qubit frequency, relaxation time $T_{1}$ and coherence time $T_{2}^{*}$. The mechanisms which limit these times are discussed, along with measurements which indicate the presence of higher transparency modes contributing to the transmon potential.

\section{Methods}

We fabricate our gatemon device from a linearly-graded Ge/SiGe heterostructure \cite{Hartmann2023}. Optical images offering a top view of the device are shown for different magnifications in Figure 1abc. A cross-sectional schematic corresponding to a longitudinal vertical cut across the semiconductor Josephson junction is shown in Figure 1d. 

The main fabrication steps are as follows. First, the top \SI{50}{\nano\meter} of the semiconductor heterostructure are removed everywhere by reactive ion etching except for an \SI{800}{\nano\meter} wide by \SI{6}{\micro\meter} long rectangular area which will form the mesa of the Josephson junction (orange section of Figure 1c). Next, we make two superconducting contacts to the Ge quantum well by etching away the top SiGe layer and then depositing Al on top of the quantum well \cite{Chotivut2024}. Two \SI{1.4}{\micro\meter} wide Al contacts \SI{300}{\nano\meter} apart define the semiconductor Josephson junction whose physical width is set by the mesa as shown in Figure 1c. The desired gatemon is obtained by shaping one of the Al contacts into a large T-shape island (blue color, Figure 1b), and shunting the other contact to the ground plane. The full device is completed with a qubit drive line (yellow), a readout resonator (purple), a reference resonator (green), and a  feedline (red), all designed to be 50-$\Omega$ and made in the same Al deposition step. Following this step, \SI{25}{\nano\meter} of Al$_{2}$O$_{3}$ is deposited by plasma-enhanced atomic layer deposition over the full chip providing the insulating gate-dielectric of the Josephson element. Finally, we fabricate a \SI{50}{\nano\meter} thick Al gate electrode over the Josephson junction (cyan) to control the carrier density in the Ge channel underneath. The gate line goes through an LC low-pass filter with a cut off frequency designed to be $\sim$ \SI{200}{\mega\hertz}.

\begin{figure}[ht]
	\centering
	 \includegraphics[width=\linewidth]{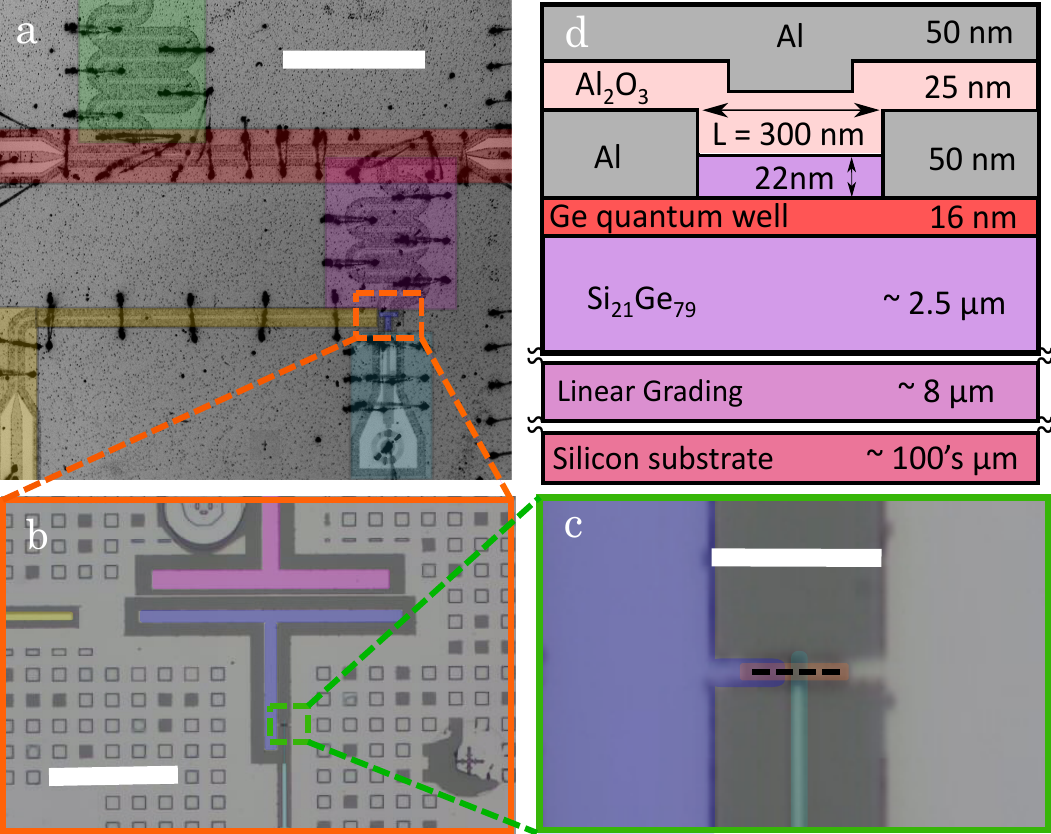}
\caption{(a) A false-color optical image of the full chip with feedline (red), reference resonator (green), readout resonator (purple), drive-line (yellow), gate layer (cyan), and charge island (blue) with a scale-bar of \SI{2}{\milli\meter}.  (b)  Zoom-in of the charge island with coupling to readout resonator and drive line with a scale-bar of \SI{100}{\micro\meter}. (c) Zoom-in of the Josephson junction showing top gate and Ge mesa (orange) with a scale bar of \SI{10}{\micro\meter}. (d) A cross-sectional schematic of the device taken at black dashed line in (c) after fabrication steps showing the superconductor/semiconductor interface with aluminum deposited on top of the germanium quantum well.}
\end{figure}

The T-shape superconducting island and the Josephson junction form a non-linear harmonic oscillator with fundamental resonance frequency $f_{01} = \sqrt{8 E_{\mathrm{C}} E_{\mathrm{J}}}/h$  where $h$ is the Planck constant, $E_{\mathrm{J}}$ is the Josephson energy, and $E_{\mathrm{C}}$ is the island charging energy \cite{Koch2007}. The charging energy is given by $E_{\mathrm{C}} = e^{2}/2\mathrm{C}$, where $e$ is the electron charge and $\mathrm{C}$ is the island capacitance to ground, which is determined by its geometry and estimated from finite-element simulations to be \SI{57}{\femto\farad} giving $E_{\mathrm{C}}/h \sim $ \SI{340}{\mega\hertz}. The Josephson energy is given by $E_{\mathrm{J}} = h I_{\mathrm{c}}/4  \pi e$ and $I_{\mathrm{c}}(V_{\mathrm{g}})$ is the critical current that depends on gate voltage ($V_{\mathrm{g}}$). From DC measurements on similar junctions (not shown), we expect the maximum critical current to be $\sim$ \SI{140}{\nano\ampere}, which is large enough to ensure the typical  transmon-regime condition $E_{\mathrm{J}} \geq 30 E_{\mathrm{C}}$. A $\lambda/2$ co-planar waveguide resonator is capacitively coupled to the measurement feedline on one hand, and to the superconducting island on the other hand, as shown in Figure 1b. Based on numerical  simulations, we estimate a qubit-resonator coupling $ g/(2\pi) \sim $ \SI{100}{\mega\hertz}. 

We measure the qubit using standard dispersive readout techniques in a dilution refrigerator with a base temperature of \SI{7}{\milli\kelvin}. We put the qubit in the dispersive regime by a large difference ($\Delta$) between the qubit frequency and the readout (RO) resonator bare frequency, i.e. $\Delta = | f_{\mathrm{q}} - f^{\mathrm{RO}}_{\mathrm{bare}}| > g/(2\pi)$. In this regime, the state of the qubit can be determined by the state of the coupled resonator from a shift ($\chi$) in its resonance frequency up to $\chi = g^{2}/\Delta$ \cite{Krantz2019}. A readout tone ($f_{\mathrm{probe}}$) is sent through the feedline to measure the resonator state and a drive tone ($f_{\mathrm{drive}}$) is sent through the drive line to induce coherent rotations of the qubit state. A schematic of the fridge wiring is shown in Supporting Information Figure S1.

\section{Results}

We begin characterizing the resonators by measuring the magnitude ($S_{21}$) and phase ($\theta$) of the transmitted radio frequency probe signal around their resonance frequency using power low enough so that the average photon number $\langle n_{ph} \rangle$ is close to 1. We fit the resonances with the same methodology as in Megrant \textit{et al.} \cite{Megrant2012} shown in Supporting Information (Figure S2). For the reference resonator, we obtain a resonance frequency $f^{\mathrm{ref}} =$ \SI{6.185}{\giga\hertz} and an internal quality factor $Q^{\mathrm{ref}}_{\mathrm{i}} \sim $ 14600. For the readout resonator, we find a higher resonance frequency, $f^{\mathrm{RO}}_{\mathrm{bare}} =$ \SI{7.799}{\giga\hertz}, and a lower internal quality factor, $Q^{\mathrm{RO}}_{\mathrm{i}} \sim$ 6800.  These values are measured while applying a large positive voltage, $V_{\mathrm{g}} =$  \SI{1.5}{\volt}, to the gate of the Josephson junction, which depletes the hole channel and hence entirely suppresses the qubit energy.

\begin{figure}[ht]
	\centering
	 \includegraphics[width=\linewidth]{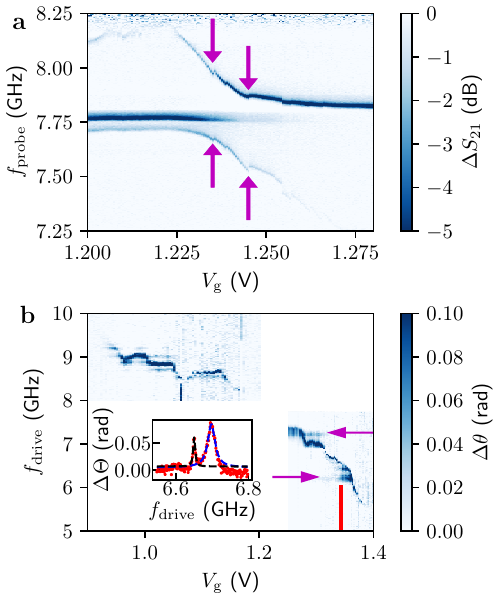}
	\caption{ (a) Resonator-qubit anti-crossing measured in transmission ($S_{21}$) as a function of probe frequency ($f_{\mathrm{probe}}$) and gate voltage ($V_{\mathrm{g}}$) showing vacuum Rabi splitting between resonator and qubit. A background is removed by subtracting an average over regions where there is no part of the anti-crossing. Third resonance line when $V_{\mathrm{g}} <$  \SI{1.25}{\volt} is not understood. Magenta arrows indicate locations of charge jumps. (b) Two-tone spectroscopy measured as a shift in resonator phase ($\theta$) as a function of drive frequency ($f_{\mathrm{drive}}$) and gate voltage showing the qubit energy can be tuned over $\sim$ \SI{3.5}{\giga\hertz}. Each vertical trace has an average subtracted. Magenta arrows indicating positions of some additional resonances. Inset shows a line-cut at the red line measured with pulsed spectroscopy with two Lorentzian fits giving the position in frequency of $f_{01}$ (blue) and $f_{02}/2$ (black).}
\end{figure}

\begin{figure*}[ht]
	 \includegraphics[width=\linewidth]{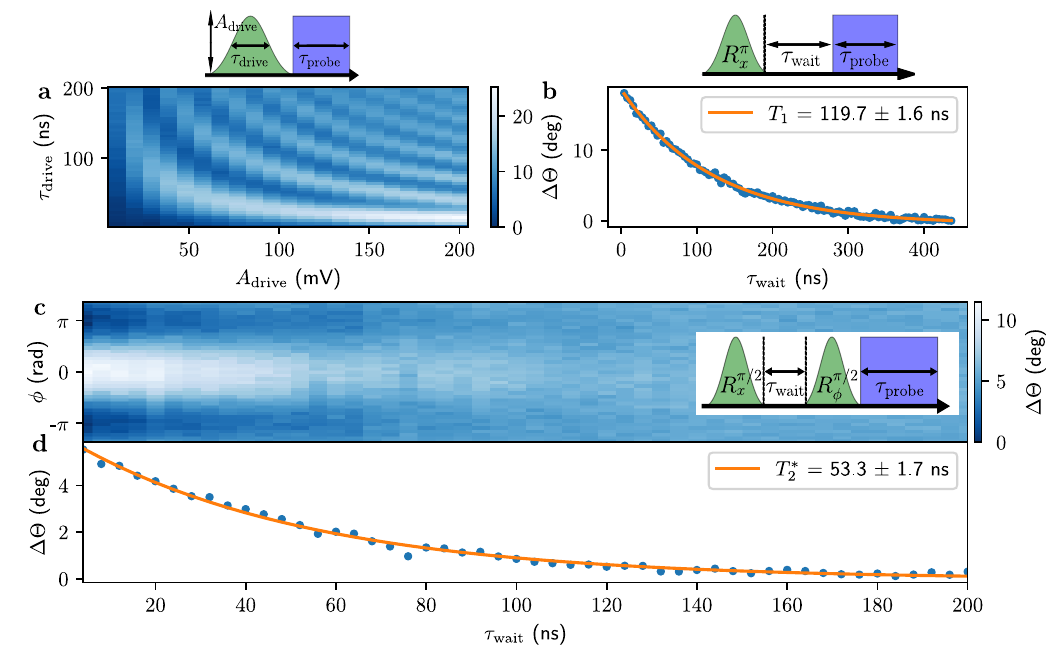}
	\caption{ Time domain measurements for $f_{\mathrm{q}} =$ \SI{6.70}{\giga\hertz}. (a) Rabi oscillations measured in the phase shift of the  readout resonator ($\Theta$) as a function of drive pulse duration ($\tau_{\mathrm{drive}}$) and amplitude ($A_{\mathrm{drive}}$), diagram of pulse sequence above, used to calibrate $\pi$ and $\pi/2$ pulses labeled $R^{\pi}$ and $R^{\pi/2}$ in (b,c,d). (b) A $T_{1}$ measurement with a wait time ($\tau_{\mathrm{wait}}$) between the $R^{\pi}$ pulse and readout pulse fit by an exponential with a diagram of pulse sequence above. (c) Ramsey measurement of two $R^{\pi/2}$ pulses where the waiting time between them ($\tau_{\mathrm{wait}}$) and the phase ($\phi$) of the second pulse $R^{\pi/2}_{\phi}$ is varied relative to the first pulse, diagram of pulse sequence inside figure.  (d) A $T_{2}^{*}$ measurement with points extracted from (c) by fitting a sine to the signal amplitude against projection angle $\phi$ for each wait time. The resulting decay in $\tau_{\mathrm{wait}}$ is fit with an exponential to obtain $T_{2}^{*}$. Each of these measurements are taken with 200,000 averages, drive pulses are Gaussian with a 4$\sigma$ cutoff, and the error in $T_{1}$ and $T_{2}^{*}$ is taken as the error of the exponential fit.}
\end{figure*}

To investigate the qubit-resonator coupling, we vary the gate voltage to tune the qubit frequency across the resonator frequency. The interaction between the qubit and electromagnetic field of the resonator results in   vacuum Rabi splitting manifesting as a level anti-crossing. This effect is shown in Figure 2a, where the transmission $S_{21}$ is plotted as a function of gate voltage and probe frequency.  We identify two anti-crossing branches whose minimum frequency difference yields a vacuum Rabi splitting $g/(2\pi) = $ \SI{146}{\mega\hertz} with the analysis shown in Supporting Information Figure S3. This value is a factor 1.5 higher than the simulated value, and the cause of this discrepancy is unclear. 

The two anti-crossing branches in Figure 2a exhibit several abrupt shifts, highlighted by vertical arrows. These shifts are likely due to charge jumps impacting the semiconductor channel of the Josephson junction, as already observed in other gatemon realizations (see, e.g., Ref. \cite{Sagi2024}).  

An additional horizontal branch can be clearly seen on the left-hand side of Figure 2a. This feature is probably associated with a spurious resonance in the circuit. It is observed for different measurement powers applied to the feedline and it seems to weakly interact with the gatemon qubit.

The qubit energy spectrum is measured by two-tone spectroscopy. A measurement of the qubit resonance frequency is shown in Figure 2b as a function of drive frequency and gate voltage at a fixed power, re-adjusting $f_{\mathrm{probe}}$ for each gate voltage. The qubit frequency decreases with $V_g$ as expected for semiconductor channels hosting hole carriers. We observe a qualitative square-root dependence of the qubit frequency, $f_{\mathrm{q}}$, which we can tune over a $\sim$ \SI{3.5}{\giga\hertz} range (see Supporting Information for more details). Additionally, we observe multiple horizontal lines, some of which are highlighted by magenta arrows. These spurious resonances exhibit weak yet visible interactions  with the qubit, manifesting as small avoided crossings.

In the inset of Figure 2b, we plot a line-cut of the two-tone data measured with pulsed spectroscopy at a fixed gate voltage indicated by a red line in Figure 2b, where the probe and the drive pulses are separated in time. Upon increasing the drive power, eventually two peaks are observed which we identify as the qubit fundamental transition $f_{01}$ and a lower frequency that we ascribe to a higher level transition in a two photon process, $f_{02}/2$. For later analysis, our methodology to extract the qubit anharmonicity ($\alpha$), is to fit each peak to a Lorentzian function giving the frequencies $f_{01}$ and $f_{02}/2$. $\alpha$ is given by $\alpha/h = 2(f_{02}/2 - f_{01})$, and at this point, we find $|\alpha|/h =$ \SI{94}{\mega\hertz}. 

Now we turn to time-domain measurements by sending pulses of a set duration, amplitude and frequency to perform coherent qubit manipulation. An example of measured Rabi oscillations for the qubit frequency at $f_{\mathrm{q}}=$ \SI{6.70}{\giga\hertz} is given in Figure 3a, with the corresponding pulse sequence diagram shown just above. The probe signal phase ($\Theta$) is recorded while varying the duration ($\tau_{\mathrm{drive}}$) and the amplitude ($A_{\mathrm{drive}}$) of the drive at the fundamental Larmor frequency $f_{01}$. As expected, the Rabi frequency increases with the drive amplitude. This measurement allows us to calibrate a $\pi$ pulse rotation labeled $R^{\pi}$ and a $\pi/2$ pulse rotation labeled $R^{\pi/2}$, which are then used in the measurements of qubit relaxation and coherence time discussed below. 

$T_{1}$ measurements are performed by introducing a wait time ($\tau_{\mathrm{wait}}$) between a $R^{\pi}$ pulse and the probe pulse. An example of these measurements is displayed in Figure 3b with the pulse sequence diagram shown just above. At $\tau_{\mathrm{wait}}=0$ the measured phase signal corresponds to the excited qubit state. Due to qubit relaxation, by increasing $\tau_{\mathrm{wait}}$, the signal decays toward the phase value associated with the qubit ground state. We obtain the qubit relaxation time by fitting this decay to an exponential of the form $A \exp(-t/T_{1}) + B$. 

$T_{2}^{*}$ measurements are performed through Ramsey interference. The corresponding pulse sequence, schematically shown in the inset of Figure 3c, consists of two consecutive $\pi/2$ pulses with a varying time delay $\tau_{\mathrm{wait}}$. The second pulse, $R^{\pi/2}_{\phi}$, is phase-shifted by $\phi$ relative to the first pulse $R^{\pi/2}$. This phase shift changes the rotation axes of the second pulse resulting in a signal oscillating with $\phi$. 

Figure 3c shows an example of Ramsey-Interference in the phase $\Delta\Theta$ of the measured signal as a function of $\tau_{\mathrm{wait}}$ and $\phi$ for a qubit frequency $f_{\mathrm{q}} =$ \SI{6.70}{\giga\hertz}. Each vertical trace is fitted to a sinusoidal function of the form $\Delta\Theta  = A \sin{(\phi + B)} + C$ whose amplitude is found to decay exponentially with $\tau_{\mathrm{wait}}$ as shown in Figure 3d. As for $T_1$, the extracted decay constant $T_{2}^{*}$ is found to be robust against small changes in the qubit frequency. 

\section{Discussion}

\begin{figure}[t]
	 \includegraphics[width=\linewidth]{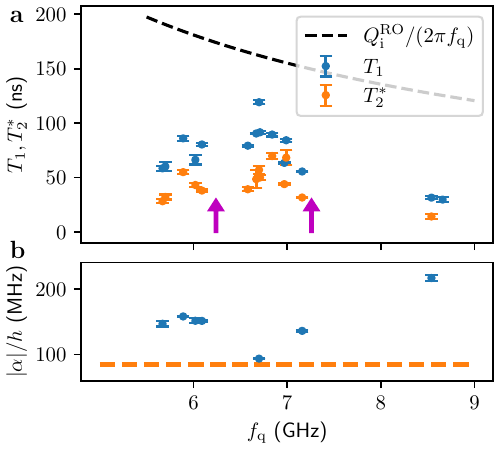}
	\caption{ (a) Extracted gatemon qubit relaxation times ($T_{1}$) and coherence times ($T_{2}^{*}$) as functions of qubit frequency ($f_{\mathrm{q}}$) with magenta arrows to indicate positions of spurious resonances found in two-tone spectroscopy measurements. The dashed black line gives the relaxation time based on the readout resonator quality factor as $ Q^{\mathrm{RO}}_{i}/(2 \pi f_{\mathrm{q}})$. (b) The qubit anharmonicity ($\alpha$) as a function of the qubit frequency extracted from two-tone spectroscopy measurements as in the inset of Figure 2b. The orange dashed line gives the lower limit of the anharmonicity as $E_{\mathrm{C}}/4$ for fully transparent modes.}
\end{figure}

The measurements of $T_{1}$ and $T_{2}^{*}$ are systematically repeated for different qubit frequencies both above and below $f^{\mathrm{RO}}_{\mathrm{bare}} $. The ensemble of results is reported in Figure 4a each taken with 200,000 averages except when $f_{\mathrm{q}} > f^{\mathrm{RO}}_{\mathrm{bare}} $ where 500,000 averages were used. Both $T_{1}$ and $T_{2}^{*}$ show no clear dependence in the explored range of qubit frequencies. However,  measurements of $T_{1}$ and $T_{2}^{*}$ do appear to be largely affected by the presence of spurious resonances, indicated by magenta arrows in Figure 4a and Figure 2b. These resonances could be associated with electromagnetic modes in the chip or in its proximal environment (we note that the sample holder used in our experiment has no electromagnetic shield \cite{Corcoles2011}). When measuring near these resonances, we clearly observe broader qubit line-widths and extra features appearing in our calibration measurements (see Supporting Information Figure S5). Moreover, these spurious resonances are particularly visible above the resonator frequency, which could explain our encountered difficulty in measuring $T_{1}$ and $T_{2}^{*}$ when $f_q > f^{\mathrm{RO}}_{\mathrm{bare}}$. 

In Figure 4a we compare the measured  gatemon relaxation times to the relaxation time of the on-chip readout resonator, as inferred from its internal quality factor, $Q^{\mathrm{RO}}_{\mathrm{i}} \sim$ 6800 (dashed line). We note that the qubit relaxation times are consistently shorter suggesting the possibility that additional  loss mechanisms may limit the qubit lifetime besides those affecting the nearby readout resonator. The quality factor of the readout resonator is certainly limited by losses originating from discontinuities in the superconducting ground plane, such as those introduced to accommodate the drive line and the gate line. These losses are likely to impact the qubit relaxation time too. They could be reduced with an improved engineering of the superconducting microwave circuit. In the best scenario, the equivalent qubit quality factor $Q^{\mathrm{qubit}}_{\mathrm{i}}=2\pi f_q T_1$ would approach the quality factor of the reference resonator, $Q^{\mathrm{ref}}_{i} = 14600$, which we expect to be mainly limited by dielectric losses from the SiGe substrate and from the Al$_{2}$O$_{3}$ cap layer covering the entire chip. Hence, the gatemon relaxation time could increase by a factor 4, but still remain relatively short ($\sim$ \SI{0.5}{\micro\s}) for practical applications. We thus conclude that a better approach would consist in fabricating the superconducting microwave circuitry on a low-loss substrate and hence connecting it to the SiGe-based Josephson element by flip-chip assembly \cite{Rosenberg2017, Hinderling2024}.

Figure 4a shows that the dephasing time $T_{2}^{*}$ is always lower than $T_{1}$. Charge noise is a probable cause of dephasing, being an ubiquitous phenomenon in semiconductors. Indeed, charge instabilities lead to the abrupt jumps clearly visible in Figure 2a. These charge rearrangements induce variations in the qubit frequency. In principle, the impact of charge noise  should be proportional to the slope of $f_q(V_g)$ \cite{Ithier2005}. Yet we find no clear correlation between $T_{2}^{*}$ and $df_q/dV_g$. Again, the interaction of the qubit with the numerous spurious modes is likely to be the main cause of the observed fluctuations in $T_{2}^{*}$. 

To determine the effect of high transparency conduction channels, we extract the anharmonicity from two-tone measurements displaying the result in Figure 4b as a function of qubit frequency. It has already been shown \cite{Kringhøj2018} that  high transparency modes change the shape of the Josephson potential which can alter the anharmonicity from $E_{\mathrm{C}}$ up to a factor of four as $|\alpha| \geq E_{\mathrm{C}}/4$ which we plot as a dashed line on Figure 4b as a lower bound. The anharmonicity we measure is, in some cases, lower than the designed $E_{\mathrm{C}}$ by a factor of three. We conclude that there is a significant contribution of high transparency channels in our junction leading to this reduction. The anharmonicity appears to have no clear trend with $f_{\mathrm{q}}$ while the values can change by a factor of two indicating a strongly gate-dependent channel transparency likely due to mesoscopic effects in the semiconductor \cite{Kringhøj2018}. 

Importantly, these anharmonicity measurements are consistent with our previous DC measurements that show a skewed current phase relation containing higher harmonics \cite{Leblanc2023,Leblanc2024}. This is logical since both of these effects are signatures of the influence of high transparency conduction channels on the Josephson potential. It is, however, particularly promising that we measure this anharmonicity change in a cQED measurement where the Josephson junction is shunted to ground with a large capacitance. Overall these results points favorably to this system being compatible for making a $\cos{(2\phi)}$ qubit since for that goal both an added shunt capacitance and a skewed current phase relation are required.

Finally, we ought to point out the differences and similarities with the closely related work by Sagi {\it et al.} \cite{Sagi2024}. The two works use Ge/SiGe heterostructures grown using different chemical-vapor-deposition (CVD) techniques and setups (our approach, based on reduced-pressure CVD on 200-mm Si(001) wafers, is closer to large-scale manufacturing). In both works, the unstrained SiGe virtual substrates are obtained by means of a forward grading approach, which consists in gradually increasing the Ge content from 0 to 79\% (here)  or  70 \% (in  Sagi {\it et al.}) at a rate of  $\sim10$\%/$\mu$m. Hence forward-graded SiGe virtual substrates consistently  show compatibility with circuit-QED architectures, as further confirmed by some other recent works reporting the fabrication of high-impedance superconducting resonators on SiGe \cite{DePalma2024,Janik2024}. All internal quality factors range between $\sim 10^3$ and $\sim 10^4$, with the bare resonator reported here showing the largest value (14800) so far.
Recently, superconducting microwave resonators with quality factors of $\sim 10^3$ could be fabricated also on reverse-graded SiGe heterostructures \cite{Nigro2024,Kang2024}.
We note that our superconducting resonators, as well as those in Sagi {\it et al.},  are covered by an ALD-deposited  Al$_2$O$_3$ layer. The internal quality factor of the readout resonator measured here (6800) is almost twice the one reported by Sagi {\it et al.} (3800). A comparable ratio is observed also between the quality factors of our bare resonator (14800) and the bare resonator reported in Valentini {\it et al.} (7000) \cite{Valentini2024}, despite the fact that the latter was not covered by Al$_2$O$_3$. This indicates that our SiGe heterostructure yields a noticeably lower level of dielectric losses, which could explain the somewhat longer gatemon relaxation times measured here. (Moreover, our gatemon design, involving a low-pass-filtered superconducting gate metal line, is likely less lossy than the one of Sagi {\it et al.}, which involves a normal-type metal  gate connected to a 50-$\Omega$ transmission line.)  For qubit frequencies between 6 and \SI{7}{\giga\hertz}, our $T_1$ values vary from 50 to \SI{120}{\nano\s},  while  $T_1 \sim $ \SI{20}{\nano\s}  in  Sagi {\it et al.}). Concerning the dephasing time, the discrepancy is somewhat lower. In the same frequency range, $T_2^*$ varies from from 35 to \SI{70}{\nano\s} in our work, and from 20 to \SI{40}{\nano\s} in Sagi {\it et. al.}. Different from our experiment, Sagi {\it et al.} were able to explore qubit frequencies as low as \SI{2.3}{\giga\hertz} where they observed an average enhancement of both the relaxation and dephasing times ($T_1 \sim T_2^{*} \sim$ \SI{70}{\nano\s}).

\section{Conclusion}

In summary, we have fabricated and characterized a gatemon qubit based on a Ge/SiGe heterostructure. We measured comparable relaxation and dephasing times ranging from a few tens to about \SI{100}{\nano\s}, likely limited by dielectric losses and other noise mechanisms possibly coming from spurious resonances associated with on-chip modes or electromagnetic standing waves in the sample  environment.
We note that the observed dephasing times are  close to those reported in Sagi {\it et al.}, and to those found in gatemons made from Ge/Si core/shell nanowires \cite{Zheng2024,Zhuo2023}.
While these characteristic times appear too short for practical applications, we foresee that circuit and setup optimization, together with the introduction of flip-chip assembly, with the superconducting microwave circuitry fabricated on a separate low-loss substrate, would remove spurious modes and reduce dielectric losses, leading to significant improvement in the qubit figures of merit. This way, the intrinsic limits on qubit relaxation and coherence time, as well as their physical origin, could be possibly unveiled, enabling an assessment of the benefits expected from the use of heterostructures embedding a buried Ge quantum with high-mobility two-dimensional holes.

\section{Author Contributions}
Device design was done by E.K. with help from V.S., E.D., and C.T. and fabricated by E.K. with processes developed from C.T. Experiments were performed by E.K. with help from L.N., C.T., B.B., S.Z., and V.S. The results were analyzed with E.K., A.L., B.B., S.Z., V.S., R.M., F.L. and S.dF. JM.H. grew the heterostructure. E.K. wrote the manuscript with help from all co-authors. The project was supervised by F.L. and S.dF.

\section{Conflict of interest disclosure}
The authors declare no competing financial interest.

\begin{acknowledgement}
This work has been supported by the PEPR ROBUSTSUPERQ (Grant No. ANR-22-PETQ-0003), ANR project SUNISIDEuP (Grant No. ANR-19-CE47-0010), the ANR project (Grant No. ANR-23-CPJ1-0033-01), PEPR PRESQUILE (Grant No. ANR-22-PETQ-0002), and the ERC starting Grant LONGSPIN (Grant No. Horizon 2020-759388). We would like to thank Georgios Katsaros for useful discussions.
\end{acknowledgement}

\section{Supporting Information Description}
In the supporting information, we provide the full fridge wiring, measurement and analysis of the resonators spectrum, spectrum and analysis of the anti-level crossing, an estimation of the long vs. short regime of the Josephson junction, an estimation of qubit energy functional dependence and finally, a comparison of two time domain measurements.

\bibliography{refs.bib}

\pagebreak
\onecolumn
\begin{center}
\textbf{\large Supporting Information: Gatemon qubit on a germanium quantum-well heterostructure}
\end{center}

\setcounter{equation}{0}
\setcounter{figure}{0}
\setcounter{table}{0}
\setcounter{page}{1}
\makeatletter

\renewcommand{\appendixname}{Supplementary Material}
\renewcommand{\thefigure}{S\arabic{figure}} \setcounter{figure}{0}
\renewcommand{\thetable}{S\arabic{table}} \setcounter{table}{0}
\renewcommand{\theequation}{S\arabic{table}} \setcounter{equation}{0}
\renewcommand{\thesection}{\Alph{section}}

\section{Fridge wiring}

A schematic of our measurement fridge is shown in Supporting Information Figure 1. We have a large attenuation on both radio frequency (RF) lines (the drive-line and probe-line) totaling to \SI{-70}{\decibel} to reduce the noise reaching the sample. This large attenuation on the drive line necessitated having a room-temperature amplifier for the pulsed spectroscopy measurements adding a gain of $\sim +$\SI{30}{\decibel}. On the returning side, we have amplifiers at the \SI{4}{\kelvin} stage and at room temperature to enhance our signal to noise ratio. The ``homemade" low pass filter on the gate-line is a series of RC and $\Pi$ filters (having a combined resistance of \SI{2.2}{\kilo\ohm}) mounted on the \SI{4}{\kelvin} stage to reduced high frequency noise at different cut off frequencies. The fridge setup was without Eco-sorb filters in the lines, electromagnetic shielding box and flux shielding box.

\begin{figure}[h!]
  		\includegraphics[width=\linewidth]{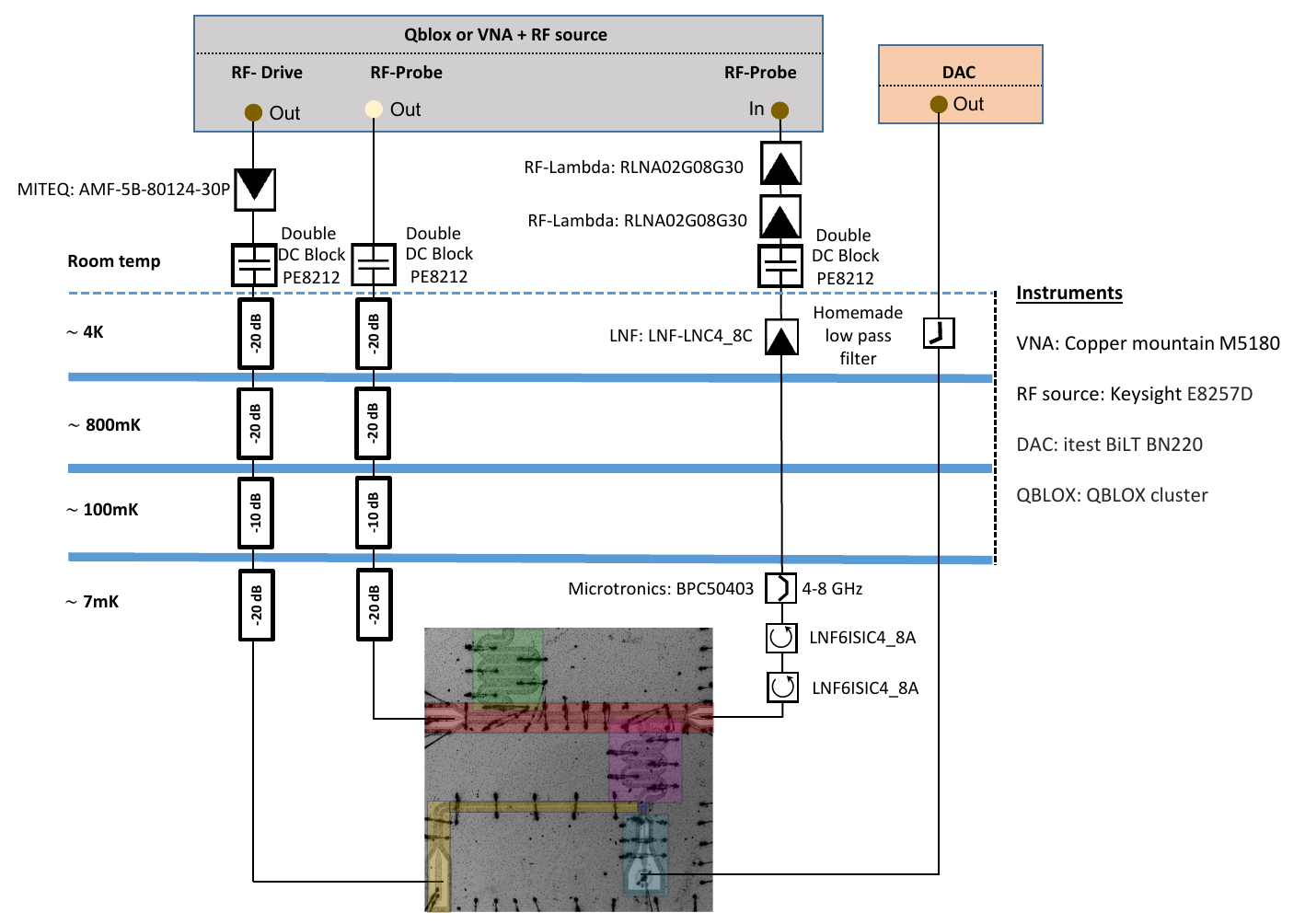}
  		\label{fig:Sfig1}
	\caption{A wiring diagram of our fridge measurement setup. The lines are either connected to i) time-domain pulsed measurement setup (QBlox) or ii) a continuous wave spectroscopy setup (VNA + RF source). }
\end{figure}

\newpage

\section{Resonator analysis}

We measure the transmission $S_{21}$ and phase as a function of probe frequency ($f_{\mathrm{probe}}$) around the resonance frequency of the readout resonator and reference resonator using a VNA with a bandwidth of \SI{10}{\hertz} with results for low photon number $<n> \sim 1$ shown in Supporting Information Figure 2ac. We measure in a frequency range around ten times the resonance full width at half maximum and adjust the background far from the resonance so that transmission is equal to \SI{0}{\decibel}.  Using the transmission $S_{21}$ and phase, measurements can be converted to the real and imaginary part of the inverse transmission shown in Supporting Information Figure 2bd. We fit the inverse transmission (red curve) with the same analysis methodology of Megrant et al. \cite{Megrant2012}, equivalent to fitting simultaneously the transmission and phase, and plot the resulting resonance dip in transmission in Supporting Information Figure 2ac (orange curve) to confirm the fit. This analysis enables us to extract the resonance frequency, the internal quality factor  ($Q_{\mathrm{i}}$) and coupling quality factor ($Q_{\mathrm{c}}$). 
\begin{figure}[h!]
  		\includegraphics[width=\linewidth]{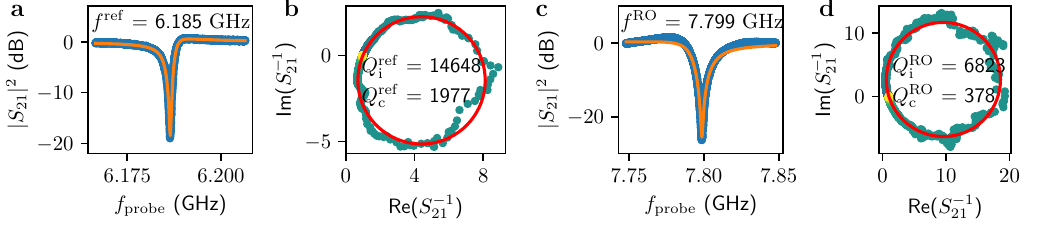}
	\caption{Measurement of resonators with analysis methodology of Megrant et al. \cite{Megrant2012} fitting in the inverse plane with $<n> \sim 1$. (ab) Fit of the reference resonator data giving $f^{\mathrm{ref}}$ = \SI{6.185}{\giga\hertz}, $Q_{\mathrm{i}}$ = 14648, and $Q_{\mathrm{c}}$ = 1977. (cd) Fit of the readout resonator, when $V_{\mathrm{g}} =$ \SI{1.5}{\volt}, giving $f^{\mathrm{RO}}_{\mathrm{bare}}$ = \SI{7.799}{\giga\hertz}, $Q_{\mathrm{i}}$ = 6823, and $Q_{\mathrm{c}}$ = 378.  }
\end{figure}

\section{Anti-crossing analysis}

We measure the anti-crossing between the resonator and gatemon resonance with a VNA probing the transmission ($S_{21}$) in a frequency range around the bare resonance frequency of the readout resonator while varying the gate voltage $V_{\mathrm{g}}$. A zoom-in of the anti-crossing measurement of main text Figure 2a, is shown in Supporting Information Figure 3a. In order to extract the frequency position of the anti-crossing upper and lower branches, we examine each vertical trace and extract the location of each dip plotting the upper, middle, and lower dip locations with an 'x' (black, blue, and red markers). The frequency difference between the upper branch and lower branch ($\Delta f$) is shown in Supporting Information Figure 3b as  a function of gate voltage. The coupling ($g/(2\pi)$), can be extracted from the Rabi vacuum splitting of the resonator with the qubit energy \cite{Larsen2015} written as $\Delta f = \sqrt{(f_{\mathrm{q}}-f^{\mathrm{RO}}_{\mathrm{bare}})^{2}+4(g/(2\pi))^{2}}$. At the minimum difference $f_{\mathrm{q}}-f^{\mathrm{RO}} \rightarrow 0$, then $\Delta f \rightarrow 2g/(2\pi)$, so we fit $\Delta f$ to a symmetric function taken to be a parabola (of the general form $A (V_{\mathrm{g}}-B)^{2}+C$) whose offset gives us an estimate for the minimum difference $\Delta f^{min} = $ \SI{292}{\mega\hertz}. Finally, this allows us to obtain an estimate for coupling of $2g/(2\pi) \sim $ \SI{146}{\mega\hertz} between the readout resonator and qubit energy which is possibly influenced by the presence of the additional middle branch marked in blue.

\begin{figure}[h!]
  		\includegraphics[width=\linewidth]{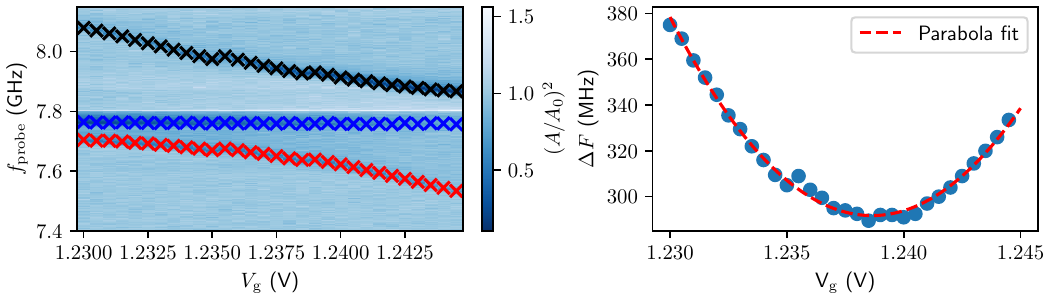}
  		\caption{The analysis to extract the $g/(2\pi)$ coupling. (a) Anti-crossing measurement with 'x' markers at each dip detection. (b) Extracted frequency difference between the upper branch (black) and lower branch (red) with a fit to a parabola to find the minimum of \SI{292}{\mega\hertz}.}
\end{figure}

\newpage

\section{Long vs short regime}
We can make an estimation of the Josephson junction regime by comparing the superconducting coherence length ($\xi$), and the mean free path ($l_{\mathrm{mfp}}$) to the length of the junction (L $\sim $ \SI{300}{\nano\meter}). In Supporting Information Figure 4a we plot the hole carrier density ($\rho_{2D}$) and mean free path ($l_{\mathrm{mfp}}$) as a function of gate voltage ($V_{\mathrm{g}}$) measured from Hall bar devices made of the same heterostructure and similar fabrication methods to aid in this analysis. Since L $ \leq l_{\mathrm{mfp}}$ for most of our measurements, we are in ballistic limit and use the formula for $\xi$ in the clean limit as $\xi = \hbar v_{f} / (\pi \Delta^{Al})$ where $v_{f}$ is the Fermi velocity, and $\Delta^{Al}$ is the superconducting gap of Aluminium \cite{Beenakker1991}. We know from DC measurements of a pure Al strip on this substrate (not shown) that the critical temperature $T^{Al}_{c}$ of the Al film is  $\sim$ \SI{1.5}{\kelvin} which, using the BCS formula $\Delta^{Al} = 1.76 k_{B} T^{Al}_{c}$  to estimate $\Delta^{Al}/e \sim $ \SI{230}{\micro\eV}.  \cite{Tinkham1996} $v_{f}$ is given by $\hbar k_{f} /m^{*} $ where $m^{*}$ is the effective mass measured to be $ \sim 10 \% $ of the electron rest mass (not shown) and the Fermi wavevector which is given from the hole density of states as $k_{f} = \sqrt{2\pi \rho_{2D}}$. Using the hole density in Supporting Information Figure 4a, we estimate $k_{f}$ to be in the range $[1.4,2] \times 10^{8}$ $m^{-1}$ giving  $\xi \sim $ [100, 200] \SI{}{\nano\meter} making our junction weakly in the long regime since $L \sim \xi$. Additionally since $l_{\mathrm{mfp}} > $ L for most of our measurements, we take it that we are in the ballistic long regime.     

\begin{figure}[h!]
  		\includegraphics[width=\linewidth]{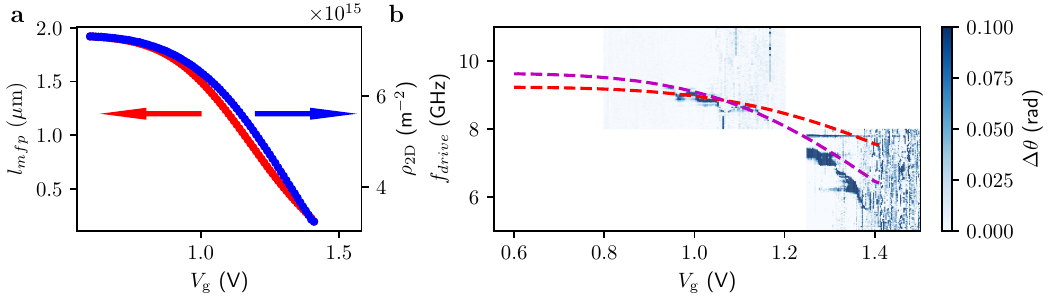}
	\caption{Estimation for the expected trend of the qubit frequency $f_{\mathrm{q}}$ with gate voltage ($V_{\mathrm{g}}$) based on (a) hole mean free path ($l_{\mathrm{mfp}}$) and hole carrier density ($\rho_{2D}$) as a function of gate voltage ($V_{\mathrm{g}}$) measured in a Hall bar device in the same heterostructure. (b) Two-tone spectroscopy measurement (same as main text Figure 2b) plotted with our expected model for $f_{\mathrm{q}} \propto \sqrt[4]{\rho_{2D}}$ (magenta curve) and $f_{\mathrm{q}} \propto \sqrt{\rho_{2D}}$ (red curve) for realistic numbers.}
\end{figure}

\section{Estimation of qubit energy functional dependence}

We can estimate the qubit energy ($f_{\mathrm{q}}$) as a function gate voltage ($V_{\mathrm{g}}$) using the expected qubit dependence on critical current ($I_{\mathrm{c}}$) and how we expect the critical current to vary based on the junction regime. As a non-linear harmonic oscillator \cite{Krantz2019}, we have $ h f_{01} = \sqrt{8 E_{\mathrm{C}} E_{\mathrm{J}}} $, with $E_{\mathrm{J}} = \hbar I_{\mathrm{c}}/2e$ and $E_{\mathrm{C}}/h \sim $ \SI{340}{\mega\hertz} from finite element simulations. Since we are in the long ballistic limit (see last section),  we take that the critical current is limited by the Thouless energy as $I_{c} = 10  G_{N} E_{\mathrm{th}}$ where $G_{N}$ is the normal state conductance and $E_{\mathrm{th}}$ is the Thouless energy in the ballistic limit given by $ \hbar v_{f}/(L \pi)$ \cite{Dubos2001}. $G_{N}$ is approximately given by the number of conducting channels multiplied by an average transparency $\tau$ as $G_{N} \sim (2W/\lambda_{f})\tau (e^{2}/h) $ where W is the width of the junction and $\lambda_{f}$ is the Fermi wavelength related to the Fermi wavevector by $\lambda_{f} = 2\pi/k_{f}$. 

Putting this altogether, we get that $ h f_{01} \propto \sqrt{\rho_{2D}} $ from one factor of $k_{f}$ from the Thouless energy and one from the normal state conductance. We plot quadratic dependence (magenta curve) using W = \SI{800}{\nano\meter}, and $\tau \sim 1\%$, over the two-tone data in Supporting Information Figure 4b, we find this general form appears to match the trend of the qubit for low gate voltages ($V_{\mathrm{g}} < $ \SI{1.3}{\volt}). If instead we had taken that we were in the short junction regime, this would remove one factor of $k_{f}$ giving $ h f_{01} \propto \sqrt[4]{\rho_{2D}} $ (red curve) which is in disagreement with the data.  At lower densities (or higher gate voltages $V_{\mathrm{g}} > $ \SI{1.3}{\volt}), it is may be that the regime changes from long ballistic (used here) to long diffusive when the mean free path $l_{\mathrm{mfp}}$ become comparable to L changing $E_{\mathrm{th}}$ and $\xi$ to a form which depends on $l_{\mathrm{mfp}}$ reducing further $f_{\mathrm{q}}$. 

\section{Comparison of two time domain measurements}

Here we compare two time domain measurements with the critical difference that one was taken in a region in frequency space far separated from resonances found in two-tone spectroscopy  $f_{\mathrm{q}} = $ \SI{6.70}{\giga\hertz} and the another $f_{\mathrm{q}} = $ \SI{7.37}{\giga\hertz} is taken in a region relatively close to a resonance at \SI{7.26}{\giga\hertz} (within $\sim \alpha/h$). Supporting Information 5 shows the calibration measurements that we take either immediately before or after $T_{1}$ and $T_{2}^{*}$ measurements to ensure that the qubit behaves as a two-level system. The measurements for $f_{\mathrm{q}} = $ \SI{6.70}{\giga\hertz}, shown in Supporting Information 5a-c, show three peaks in two-tone spectroscopy (Supporting Information 5a) that correspond to frequencies of coherent oscillations in Rabi chevrons measurement (Supporting Information 5b) with lower Rabi frequencies for the higher order transitions ($f_{02}/2$ or $f_{03}/3$). Additionally, the Ramsey measurement (Supporting Information 5c) appears to have only one relevant frequency. 

In contrast, the two-tone measurement of $f_{\mathrm{q}} = $ \SI{7.37}{\giga\hertz} in Supporting Information 5d has three peaks (although the peak at \SI{7.26}{\giga\hertz} did not move with gate voltage), but the corresponding Rabi chevrons in Supporting Information 5e show the opposite trend where higher Rabi frequencies are shown for the higher order transitions. Additionally, the Ramsey measurement in Supporting Information 5f appears to be influenced by many frequencies. It was these differences in calibration measurements that made us exclude these regions from our results appearing in the main text Figure 4a.  

\begin{figure}[h!]
  		\includegraphics[width=\linewidth]{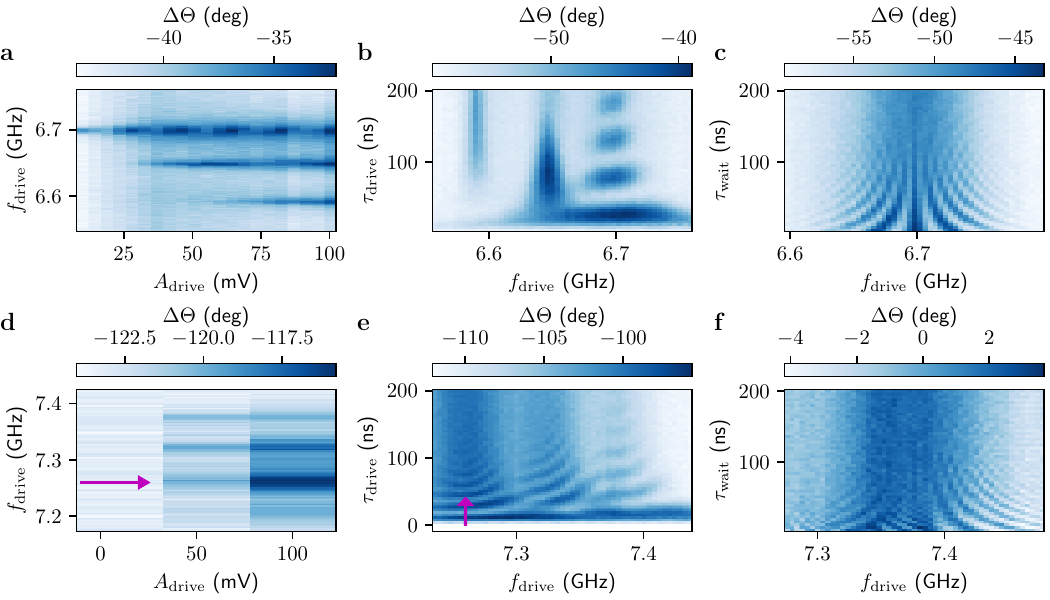}
	\caption{Additional qubit calibration measurements for (a-c) $f_{01} = $ \SI{6.70}{\giga\hertz} (d-f) $f_{01} = $ \SI{7.37}{\giga\hertz}. (ad) Two-tone spectroscopy measurements as a function of $f_{\mathrm{drive}}$ and amplitude of sent power $A_{\mathrm{drive}}$. (be) Rabi chevrons measurement over the same $f_{\mathrm{drive}}$ range as in (a) as a function of pulse duration  $\tau_{\mathrm{drive}}$. (cf) Ramsey measurement of two $\pi/2$ pulses as a function of $f_{\mathrm{drive}}$ and wait time $\tau_{\mathrm{wait}}$ between $\pi/2$ pulses. Magenta arrow highlight resonance at \SI{7.26}{\giga\hertz} the same position as shown in main text Figure 2(b).}
\end{figure}

\end{document}